\documentclass[aps,prl,twocolumn,noshowpacs]{revtex4-2}
\usepackage{amsmath}
\usepackage{graphicx}
\usepackage{amsbsy}
\usepackage{amssymb}
\usepackage{amsmath}
\usepackage{siunitx}
\usepackage{epsfig}
\usepackage{latexsym}
\usepackage{wasysym}
\usepackage{pifont}
\usepackage{mathrsfs}
\usepackage{natbib}
\usepackage{float}
\newfloat{widefig}{thp}{lop}
\usepackage{comment}
\usepackage{verbatim}
\usepackage{multirow,array}
\usepackage{hyperref}
\usepackage{braket}
\usepackage{xcolor}
\usepackage{romannum}
\usepackage{soul}

\renewcommand{\vec}[1]{\mathbf{#1}}

\begin{document}

\title{Electron-induced non-monotonic pressure dependence of the lattice thermal conductivity of $\theta$-TaN}

\author{Ashis Kundu$^{2,3}$}
\author{Yani Chen$^{2,1}$}
\author{Xiaolong Yang$^{4}$}
\author{Fanchen Meng$^{5,6}$}
\author{Jes\'us Carrete$^{7,8}$}
\author{Mukul Kabir$^{3}$}
\author{Georg K. H. Madsen$^{8}$}
\author{Wu Li$^{1,2}$}
\email[]{wu.li.phys2011@gmail.com}
\affiliation{$^{1}$Eastern Institute for Advanced Study, Eastern Institute of Technology, Ningbo 315200, China}
\affiliation{$^{2}$Institute for Advanced Study, Shenzhen University, Shenzhen 518060, China}
\affiliation{$^{3}$Department of Physics, Indian Institute of Science Education and Research (IISER) Pune, P.O. 411008, India}
\affiliation{$^{4}$College of Physics and Center of Quantum Materials and Devices, Chongqing University, Chongqing 401331, China}
\affiliation{$^{5}$Research Computing and Data, Clemson University, Clemson, SC, 29634, USA}
\affiliation{$^{6}$Center for Functional Nanomaterials, Brookhaven National Laboratory, Upton, New York 11973, USA}
\affiliation{$^{7}$Instituto de Nanociencia y Materiales de Aragón (INMA), CSIC-Universidad de Zaragoza, E-50009 Zaragoza, Spain}
\affiliation{$^{8}$Institute of Materials Chemistry, TU Wien, A-1060 Vienna, Austria}

\begin{abstract}

Recent theoretical and experimental research suggests that $\theta$-TaN is a semimetal with high thermal conductivity ($\kappa$), primarily due to the contribution of phonons ($\kappa_\texttt{ph}$). By using first-principles calculations, we show a non-monotonic pressure dependence of the $\kappa$ of $\theta$-TaN. $\kappa_\texttt{ph}$ first increases until it reaches a maximum at around 60~GPa, and then decreases. This anomalous behaviour is a consequence of the competing pressure responses of phonon-phonon and phonon-electron interactions, in contrast to the known materials BAs and BP, where the non-monotonic pressure dependence is caused by the interplay between different phonon-phonon scattering channels. Although TaN has phonon dispersion features similar to BAs at ambient pressure, its response to pressure is different and an overall stiffening of the phonon branches takes place. Consequently, the relevant phonon-phonon scattering weakens as pressure increases. However, the increased electronic density of states near the Fermi level, and specifically the emergence of additional pockets of the Fermi surface at the high-symmetry L point in the Brillouin zone, leads to a substantial increase in phonon-electron scattering at high pressures, driving a decrease in $\kappa_{\mathrm{ph}}$. At intermediate pressures ($\sim$~20$-$70~GPa), the $\kappa$ of TaN surpasses that of BAs. Our work provides deeper insight into phonon transport in semimetals and metals where phonon-electron scattering is relevant.

\end{abstract}
\maketitle
\pagenumbering{arabic}


The pressure response of materials is critical to understanding their properties and performance in many areas, including condensed-matter physics, materials science, and geophysics~\cite{Zhang_NRM17,Miao_NRC20,Knudson_SC15,Fratanduono_SC21,Pozzo_N12,Konopkova_N16,Zhou_NRP22}. Among the many properties affected by pressure, the thermal conductivity ($\kappa$) of a material stands out as a fundamental transport parameter for heat conduction. $\kappa$ is limited by resistive scattering between intrinsic carriers such as phonons and electrons~\cite{Ziman_book01}. Modulation of $\kappa$ with pressure and temperature over a wide range can help in understanding heat conduction in the Earth's interior~\cite{Pozzo_N12,Konopkova_N16} and developing mechanisms to control thermal management in microelectronic devices~\cite{Ball_NN12,Garimella_IEEE12,Waldrop_NN16}.

Under hydrostatic pressure, a crystal’s volume decreases, generally resulting in a hardening of the phonon modes. This typically leads to an increase in lattice $\kappa$ unless a structural phase transition occurs~\cite{Bridgman_RMP35,Ziman_book01}. This intuition is supported by various model calculations, although those have also shown that a large mass ratio can reverse the trend~\cite{Lindsay_PRB15}. In fact, a recent ab-initio study has revealed a non-monotonic pressure dependence of $\kappa$ for BAs, BSb and BP, with BAs and BSb share a similar microscopic origin underlying this behavior, which differs from that of BP~\cite{Ravichandran_NC19,Ravichandran_NC21}. Experimental observations for BAs also support this finding~\cite{Li_N22}.


In our previous work, we predicted that the metallic compound TaN has an ultrahigh thermal conductivity, mainly contributed by phonons and higher than that of the best electrical conductor, silver. This was found to be due to its unique combination of phononic, electronic, and nuclear properties~\cite{Kundu_PRL21}. Among the transition-metal carbides and nitrides, TaN was found to be the only compound with such high $\kappa$. Our research also revealed that the mechanism responsible for TaN's ultrahigh $\kappa$ is similar to that of BAs~\cite{Feng_PRB17}, with an additional factor of weak phonon-electron scattering. Recent experiments have indicated that this predicted high ambient-pressure value of $\kappa$ for TaN can be achieved by increasing the grain size of samples and decreasing their defect concentrations~\cite{Lee_AFM23}.


In this study, we investigate the impact of pressure on $\kappa$ in TaN using first-principles calculations. Our findings reveal a non-monotonic pressure dependence of $\kappa$ due to competing responses of phonon-phonon and phonon-electron interactions. However, the hardening of all phonon modes and the decrease in phonon scattering strength with increased pressure indicate that the non-monotonic behavior arises from strong phonon-electron scattering at higher pressure, which is influenced by the increased electronic density of states around the Fermi energy.  Thus our work provides a new perspective on the non-monotonic pressure dependence of $\kappa$ in TaN.

In the framework of the linearized Boltzmann transport equation (BTE), the lattice thermal conductivity tensor ($\kappa_\texttt{ph}^{\alpha \beta}$) is expressed as~\cite{ShengBTE}

\begin{equation}
    \kappa_\texttt{ph}^{\alpha \beta} = \frac{1}{8\pi^3}\int\limits_{\mathrm{BZ}} \sum\limits_p C_V\left(p\vec{q}\right)v^\alpha_{p\vec{q}} F^\beta_{p\vec{q}}d^3\vec{q},
    \label{eq:sigma}
\end{equation}

\noindent where $p$ runs over all phonon branches, BZ denotes the Brillouin zone of the crystal, $\alpha$ and $\beta$ are Cartesian axes, and $\omega_{p\vec{q}}$, $C_V\left(p\vec{q}\right)$, $v^\alpha_{p\vec{q}}$,  and $F^\beta_{p\vec{q}}$ are the angular frequency, mode heat capacity at constant volume, group velocity and mean free displacement, respectively, of phonons from branch $p$ with wave vector $\vec{q}$.

Phonon scattering is the main ingredient for determining $\kappa_{\mathrm{ph}}$. It enters the expression above through $F^\beta_{p\vec{q}}$. $F^\beta_{p\vec{q}}$ is determined by~\cite{ShengBTE}

\begin{equation}
    F^\beta_{p\vec{q}} = \tau_{p\vec{q}} \left(v^{\beta}_{p\vec{q}}+\Delta^{\beta}_{p\vec{q}}\right).
    \label{eq:fpq}
\end{equation}

\noindent Here, $\tau_{p\vec{q}}$ is the phonon lifetime and $\Delta^{\beta}_{p\vec{q}}$ depends linearly on $F^\beta_{p\vec{q}}$ and describes the deviation from the relaxation-time approximation (RTA). We obtain $F^\beta_{p\vec{q}}$ iteratively starting from the RTA~\cite{ShengBTE,Li_PRB15}.

The inverse of $\tau_{p\vec{q}}$ is the scattering rate and is expressed as a sum of the contributions from the different scattering mechanisms. We include all the essential scattering mechanisms to calculate the thermal conductivity. Specifically, we take three-phonon (3ph), four-phonon (4ph), phonon-isotope (ph-iso), and phonon-electron (ph-el) scattering into account in our calculations.
\begin{equation}
    \frac{1}{\tau_{p\vec{q}}} = \frac{1}{\tau^{\mathrm{3ph}}_{p\vec{q}}} + \frac{1}{\tau^{\mathrm{4ph}}_{p\vec{q}}} + \frac{1}{\tau^{\mathrm{ph\mbox{-}iso}}_{p\vec{q}}} + \frac{1}{\tau^{\mathrm{ph\mbox{-}el}}_{p\vec{q}}}.
    \label{eq:scattering}
\end{equation}
\noindent   Phonon-electron scattering involves only one phonon and does not affect $\Delta^{\beta}_{p\vec{q}}$~\cite{Chen_PRB19}. Furthermore, four-phonon scattering is dominated by Umklapp processes\cite{Feng_PRB17,Yang_PRB19} and their contribution to $\Delta^{\beta}_{p\vec{q}}$ can be neglected. The detailed expressions for $\Delta_{p\vec{q}}$, $1/\tau^{\mathrm{ph\mbox{-}iso}}_{p\vec{q}}$, $1/\tau^{\mathrm{3ph}}_{p\vec{q}}$, $1/\tau^{\mathrm{4ph}}_{p\vec{q}}$, $1/\tau^{\mathrm{ph\mbox{-}el}}_{p\vec{q}}$ can be found in Refs.~\citenum{Ward_PRB09,ShengBTE,Li_PRB15,Liao_PRL15,Chen_PRB19,Feng_PRB16,Feng_PRB17}.

To solve the phonon BTE, we need harmonic (second-order) and anharmonic (third- and fourth-order) interatomic force constants (IFCs), which are extracted from density-functional-theory calculations carried out using the projector-augmented plane wave method~\cite{PAW94} as implemented in VASP~\cite{VASP196,VASP299}. For the calculation of phonon-electron scattering, we used the EPW software~\cite{EPW_16}, based on maximally localized Wannier functions, along with the \textsc{quantum espresso} package~\cite{QM_09}. We used the FourPhonon package~\cite{FourPhonon} to compute the four-phonon scattering rates. Finally, we used a modified ShengBTE package~\cite{ShengBTE} for calculating $\kappa_{\mathrm{ph}}$ after including phonon-electron and four-phonon scattering rates. The temperature induced phonon renormalization effect is considered using self-consistent phonon (SCPH) theory~\cite{Errea_PRB14,Eriksson_ATS19}, and it is found to have a minimal impact on $\kappa_{\mathrm{ph}}$ (see Supplemental Material~\cite{supp}). The coherence effect of the phonons~\cite{Simoncelli_NP19,Isaeva_NC19,Fiorentino_PRB23,Simoncelli_NCM23} is not considered in our calculations as we estimate that at ambient pressure the average phonon interband spacing ($\sim 1.5-2$ THz) in TaN is orders of magnitude larger than the phonon linewidths ($\sim 0.001-0.1$ THz) and can be considered negligible. Further computational details, including our approach to calculating the electronic contribution to thermal transport, are given in the Supplemental Material~\cite{supp}.


\begin{figure}[t]
\centerline{\hfill
    \includegraphics[width=0.40\textwidth]{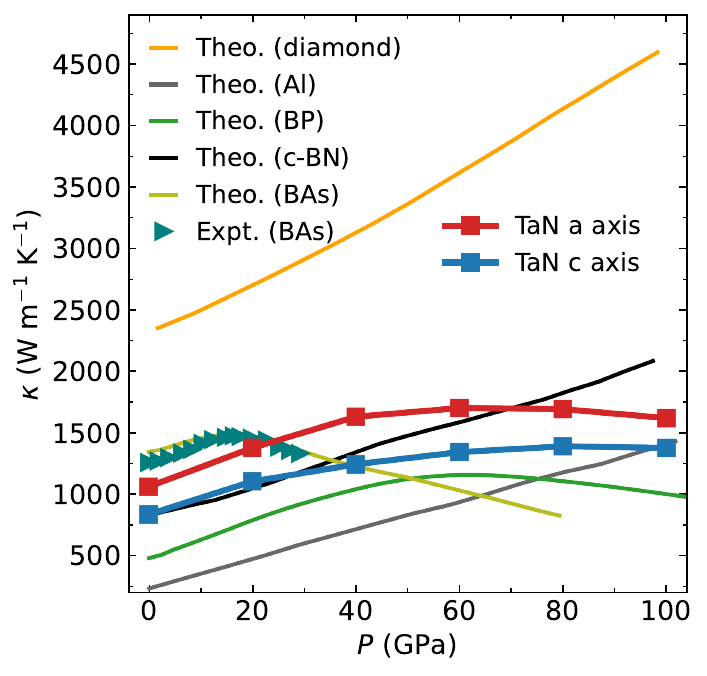}\hfill}
\caption{Calculated room-temperature total thermal conductivity ($\kappa=\kappa_\texttt{ph} + \kappa_\texttt{e}$) as a function of pressure for TaN along the a and c axes in comparison with available results of $\kappa$ under pressure. Theoretical results of $\kappa$ for diamond~\cite{Broido_PRB12}, Al~\cite{Giri_JPCL22}, BP~\cite{Ravichandran_NC21}, c-BN~\cite{Ravichandran_NC19}, and BAs~\cite{Ravichandran_NC19} and experimental results for BAs (right-pointing triangles~\cite{Li_N22}) under pressure are also shown.}
\label{fig1}
\end{figure}

Figure~\ref{fig1} shows the calculated pressure dependence of the total thermal conductivity ($\kappa$) of TaN compared to diamond~\cite{Broido_PRB12}, BAs~\cite{Ravichandran_NC19,Li_N22}, c-BN~\cite{Ravichandran_NC19}, BP~\cite{Ravichandran_NC21}, and Al~\cite{Giri_JPCL22}, which have been reported in the literature. At ambient pressure, the room-temperature total $\kappa$ of TaN is 1037 and 824 $\mathrm{W\,m^{-1}\,K^{-1}}$ along the a and c axes, respectively, consistent with the values reported earlier~\cite{Kundu_PRL21,Lee_AFM23}. The electronic contributions to the thermal conductivity ($\kappa_\texttt{el}$) at ambient pressure are 27 and 15 $\mathrm{W\,m^{-1}\,K^{-1}}$, respectively. They remain negligible at high pressures (see Supplemental Material~\cite{supp}). Interestingly, the non-monotonic behavior revealed originally in BAs~\cite{Ravichandran_NC19,Li_N22} and BP~\cite{Ravichandran_NC21} also occurs in TaN.  However, that non-monotonic behavior diminishes at high temperatures. Above 20~GPa, TaN overtakes BAs as the bulk material with the second largest $\kappa$. In contrast, $\kappa$ increases monotonically in diamond~\cite{Broido_PRB12}, c-BN~\cite{Ravichandran_NC19}, and Al~\cite{Giri_JPCL22}. $\kappa$ is dominated almost exclusively by the electronic contribution in metals. The increase of $\kappa$ in Al is mainly attributable to the decrease of phonon-electron coupling. In usual cases like diamond and c-BN, the phonon branches stiffen, and the three-phonon scattering weakens under pressure. A monotonic increase in $\kappa$ with pressure follows. In BP, a gap between acoustic and optical modes ($a$-$o$ gap) is present in the phonon dispersion at ambient pressure. The widening of the $a$-$o$ gap with rising pressure decreases the scattering involving two acoustic and one optical phonon ($aao$), driving $\kappa$ to increase at low pressures. In BAs, the $a$-$o$ gap is so large that $aoo$ scattering is irrelevant. However, four-phonon scattering plays an important role in $\kappa$. The thermal conductivity increases with increasing pressures in the low-pressure regime because of the progressive weakening of four-phonon scattering. In both BP and BAs, the reduced bunching (the proximity of acoustic phonon branches) in the phonon dispersion with increasing pressure, leading to enhanced scattering among three acoustic phonons ($aaa$ processes), is the origin of the decreasing $\kappa$ at high pressure.

\begin{figure}[t]
	\centerline{\hfill
		\includegraphics[width=0.48\textwidth]{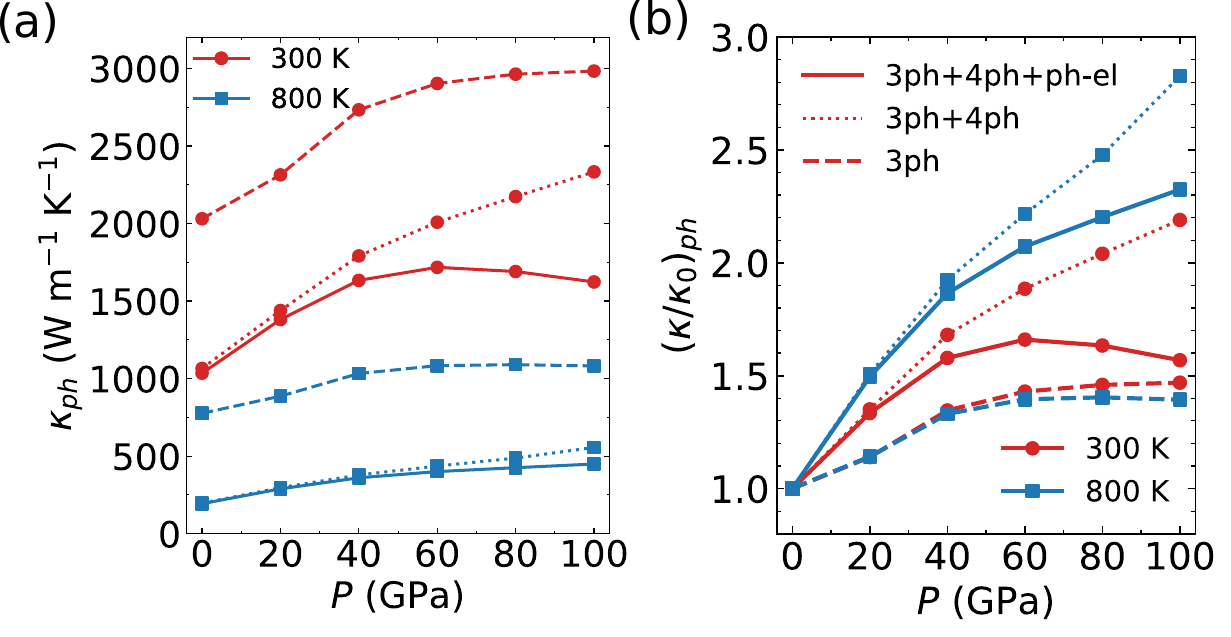}\hfill}
	\caption{(a) Calculated lattice thermal conductivity ($\kappa_\texttt{ph}$) and (b) scaled $\kappa_\texttt{ph}$ for TaN along the a axis for different pressures with different combinations of scattering mechanisms at 300 and 800~K.} 
	\label{fig2}
\end{figure}

TaN has a phonon dispersion similar to BAs as regards the presence of a large $a$-$o$ gap and the bunching of acoustic phonons, a key characteristic leading to high $\kappa$ in TaN and BAs~\cite{Kundu_PRL21}. To see whether the mechanism underlying the non-monotonic change of $\kappa$ in TaN is the same as in BAs, we show the pressure-dependent $\kappa_\texttt{ph}$ calculated by considering only specific combinations of phonon scattering mechanisms for two different temperatures, 300 and 800~K, in Fig.~\ref{fig2} (a). Surprisingly, when only considering three phonon scattering, $\kappa_\texttt{ph}$ constantly increases with pressure in TaN, in contrast to BAs. Similar to BAs, the four-phonon scattering plays a vital role in determining $\kappa_\texttt{ph}$ in TaN. At ambient pressure, $\kappa_\texttt{ph}$ is reduced by 48\% after including four-phonon scattering on top of three-phonon scattering. Although four-phonon scattering has a reduced effect on $\kappa_\texttt{ph}$ at higher pressures, it can still lead to a 27\% reduction in $\kappa_\texttt{ph}$ at 80~GPa as compared to the case with three-phonon scattering only. By comparison, in BAs the effect of four-phonon scattering is weak at higher pressure (e.g. 17\% at 51~GPa). When further including phonon-electron scattering, the influence on $\kappa_\texttt{ph}$ begins to manifest at 40~GPa. Above 60~GPa, $\kappa_\texttt{ph}$ can be significantly reduced by phonon-electron scattering, which is so strong as to cause $\kappa_\texttt{ph}$ to decrease with pressure. Therefore, the non-monotonic pressure dependence of $\kappa_\texttt{ph}$ is a consequence of the interplay between the phonon-phonon and phonon-electron scattering, rather than between three-phonon and four-phonon scattering channels as in BAs~\cite{Ravichandran_NC19}, and the one between $aaa$ and $aao$ scattering channels as in BP~\cite{Ravichandran_NC21}. Next, we will elucidate why $\kappa_\texttt{ph}^{(3)}$ in TaN behaves differently from BAs and why phonon-electron scattering gives rise to decreasing  $\kappa_\texttt{ph}$ at high pressures.
 
\begin{figure}[t]
	\centerline{\hfill
		\includegraphics[width=0.46\textwidth]{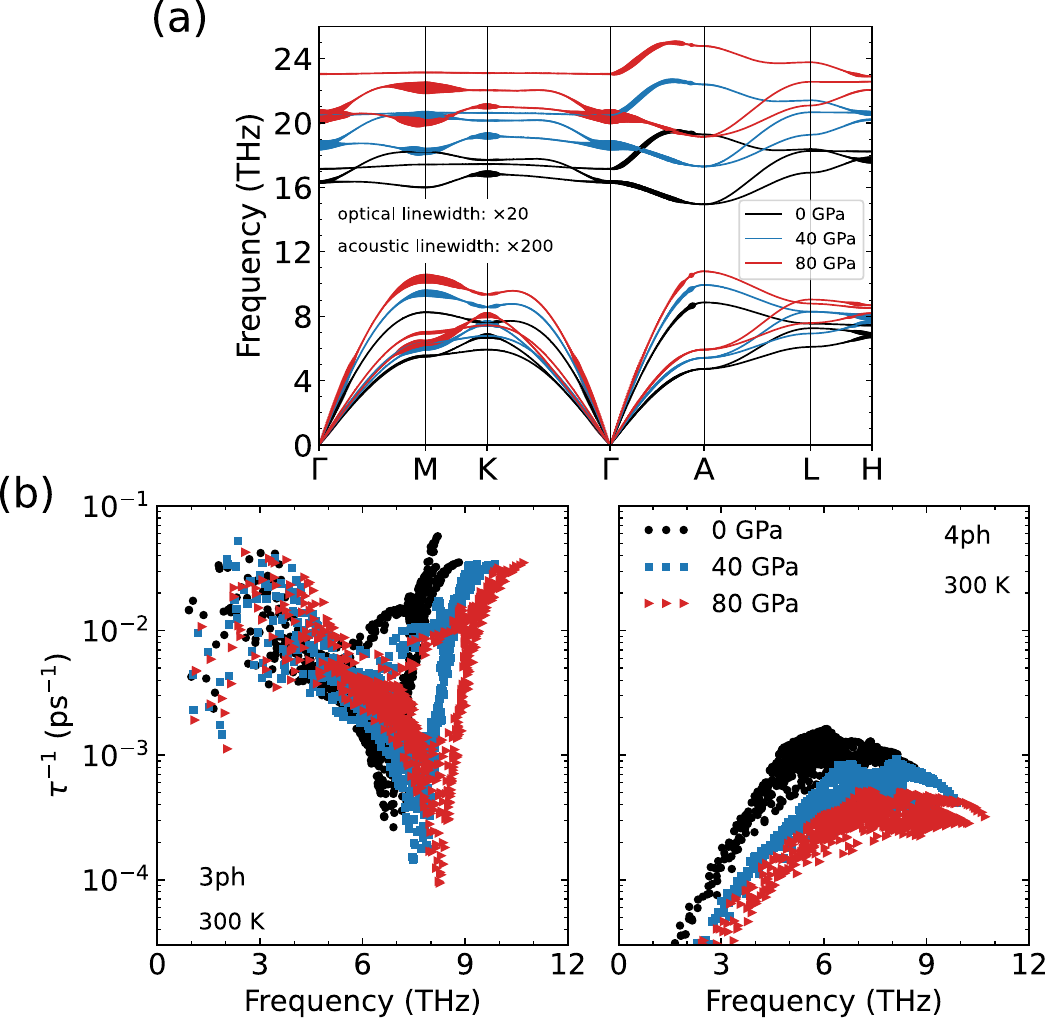}\hfill}
	\caption{(a) Calculated phonon dispersion along high-symmetry directions for TaN for different pressures.The size of circles superimposed onto the line corresponds to phonon linewidth due to phonon-electron interaction. The linewidths of acoustic and optical phonons are scaled by factors of 200 and 20, respectively. (b) Calculated three-phonon (3ph) and four-phonon (4ph) scattering rates for TaN for different pressures at 300~K.}
	\label{fig3}
\end{figure}


Due to a large $a$-$o$ gap combined with the small bandwidth of optical phonon frequencies, the three-phonon scattering of the heat-carrying acoustic phonons is limited by $aaa$ processes in BAs~\cite{Ravichandran_NC19}. Further, due to bunching, the $aaa$ scattering rates are extremely weak, with a characteristic dip in the intermediate frequency range. With increasing pressure, the longitudinal acoustic phonons stiffen evidently while the transverse acoustic phonons change marginally. Therefore, the bunching effect is significantly reduced, increasing the scattering phase space and causing the $aaa$ scattering rates to increase fast.

In the case of TaN, the phonon dispersion and three-phonon scattering rates for different pressures are plotted in Fig.~\ref{fig3} (a) and (b), respectively. The $a$-$o$ gap is not large enough to completely suppress the $aao$ scattering at high acoustic frequencies, and the $aao$ scattering is reduced partially due to a larger $a$-$o$ gap under pressure (see Supplemental Material~\cite{supp}). The optical bandwidth even allows $aoo$ processes to dominate the scattering rates at low frequencies, and the $aoo$ scattering is strengthened partially because of a larger optical band width under pressure. However, the primary contribution to $\kappa_\texttt{ph}^{(3)}$ arises from acoustic phonons in an intermediate frequency range, for which scattering is almost exclusively limited by $aaa$ processes. Under applied pressure, all acoustic phonon modes experience hardening at the same pace, as manifested by the fact that the acoustic phonon dispersions can overlap if scaled properly [Fig.~\ref{fig3} (a)](see Supplemental Material~\cite{supp}). This indicates that the bunching of acoustic modes remains nearly unchanged with increasing pressure in TaN. The difference in the response of the phonon dispersion, particularly the acoustic part to pressure, gives rise to different responses of the scattering rates to pressure. The $aaa$ scattering rate for a given $\textbf{q}$ point decreases with pressure in TaN (see Supplemental Material~\cite{supp}) compared to the increasing behavior in BAs. As a consequence, $\kappa_\texttt{ph}^{(3)}$ monotonically increases with pressure in TaN.  

As shown in Fig.~\ref{fig2} (a), the four-phonon scattering plays an important role in determining $\kappa_\texttt{ph}$ in TaN throughout the whole pressure range considered, and the four-phonon effect becomes weaker at higher pressure. To understand this, the four-phonon scattering rates are plotted in Fig.~\ref{fig3} (b). Similarly to the three-phonon scattering, especially for the $aaa$ processes, the four-phonon scattering rates decrease with increasing pressure but more rapidly. This stronger pressure dependence of four-phonon scattering leads to a weaker effect at higher pressure. Consequently, the actual $\kappa_\texttt{ph}$ has a stronger pressure dependence than it would have if only three-phonon scattering was considered. This is more apparent when plotting the pressure-dependent $\kappa_\texttt{ph}$ scaled by the corresponding zero-pressure value with specific combinations of phonon scattering mechanisms considered [Fig.~\ref{fig2} (b)]. As a matter of fact, $\mathrm{d}\kappa/\mathrm{d}P$ for TaN is 17.3 and 14.2 $\mathrm{W\,m^{-1}\,K^{-1}\,GPa^{-1}}$ along the a and c axes, respectively, which are similar to the values of approximately 19.6, 15.4, and 12.6 $\mathrm{W\,m^{-1}\,K^{-1}\,GPa^{-1}}$ for diamond~\cite{Broido_PRB12}, BP~\cite{Ravichandran_NC21}, and BAs~\cite{Ravichandran_NC19}, respectively. In addition to the stronger pressure dependence, the four-phonon scattering also has a stronger $T$ dependence than the three-phonon scattering. Therefore, the actual $\kappa_\texttt{ph}$ has a stronger pressure dependence at higher temperatures for BAs. In contrast, the pressure dependence of $\kappa_\texttt{ph}^{(3)}$ for TaN is almost $T$ independent [Fig.~\ref{fig2} (b)].


\begin{figure}[t]
\centerline{\hfill
    \includegraphics[width=0.42\textwidth]{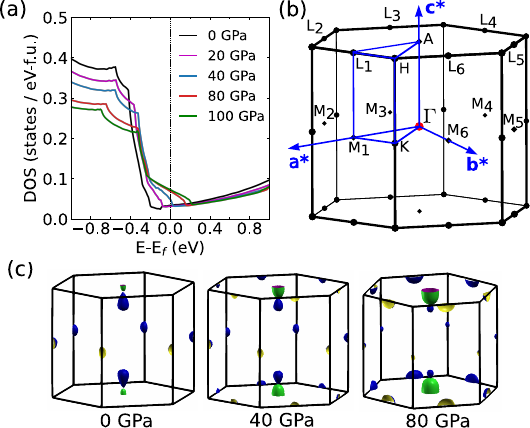}\hfill}
\caption{(a) Calculated electronic DOS for TaN for different pressures. E$_{f}$ denotes the Fermi energy. (b) First Brillouin zone of TaN with high-symmetry points labeled. (c) Calculated Fermi surfaces for TaN for different pressures.}
\label{fig4}
\end{figure}


The fact that four-phonon scattering has a stronger pressure dependence than three-phonon scattering can be understood from the weighted phase space\cite{ShengBTE,Li_PRB14,Li_PRB14Ir}, which is extremely sensitive to changes in the phonon spectrum. For example, if the entire spectrum is scaled by a factor of $c$, then the three-phonon weighted phase space is scaled by a factor of $c^{-5}$, while the four-phonon weighted phase space can be scaled by $c^{-6}$. At 80~GPa, the acoustic phonon frequency increased by a factor of $1.215$ compared to the frequency at 0~GPa. The ratios $1.215^{5} = 2.70$ and $1.215^{6} = 3.30$ closely correspond to the obtained weighted phase space ratios of $2.5$ and $3.6$ for three-phonon and four-phonon, respectively, between 0 and 80~GPa. Hence, the higher weighted phase-space ratio suggests that four-phonon scattering weakens faster than three-phonon scattering with increasing pressure (see Supplemental Material~\cite{supp}).


Finally, we study the phonon-electron scattering, leading to decreasing  $\kappa_\texttt{ph}$ at high pressures. At ambient pressure, phonon-electron scattering is negligible compared to anharmonic phonon-phonon scattering. This weak phonon-electron scattering is actually a key factor resulting in the high value of $\kappa$ for TaN compared to WC with the same type of crystalline structure, as established in our previous work~\cite{Kundu_MTP20,Kundu_PRL21}. Phonon-electron scattering is positively correlated with the electronic density of states (DOS) at the Fermi level~\cite{Xu_PRL14,Wang_JAP16,Dongre_JMCA20,Kundu_MTP20}. The weak phonon-electron scattering is attributed to the direct $d-d$ bonding between Ta atoms, resulting in a semimetallic band structure and a low DOS around the Fermi energy~\cite{Wijeyesekera_IC84}. However, the phonon-electron scattering rates can be increased as pressure increases (see Supplemental Material~\cite{supp}). Fig.~\ref{fig4} (a) illustrates that there is a simultaneous increase in the DOS at the Fermi level with increasing pressure. Consequently, the DOS around the Fermi energy gradually increases and starts to become significant at 40~GPa. Therefore, the increased DOS and subsequently higher phonon-electron scattering rates are at the origin of the decrease of $\kappa_\texttt{ph}$ at high pressures in TaN. Note that the phonon-electron scattering rates are generally insignificant even in metals with large DOS at the Fermi level~\cite{Liao_PRL15,Wang_JAP16,Jain_PRB16}. Only a few expectations have been reported recently in systems with weak anharmonic scattering, including transition metal carbides~\cite{Li_PRL18,Kundu_MTP20} and tungsten~\cite{Chen_PRB19}.

To understand the increase in DOS around the Fermi energy and its effect on phonon-electron scattering, we examine the changes in Fermi surfaces and the mode projected phonon linewidth~\cite{Chen_NPJCM19} due to phonon-electron interaction with increasing pressure, shown in Fig.~\ref{fig4} (c) and Fig.~\ref{fig3} (a), respectively. The phonon-electron scattering rates are proportional to the phonon linewidth through a factor of $2/\hbar$~\cite{Liao_PRL15}. As pressure rises, additional pockets of the Fermi surface appear at L point, in line with band structure's evolution where the band around the L point intersects the Fermi level due to increased Ta-$d$ ($d_{xy}+d_{x2-y2}$) and N-$p$ ($p_{z}$) hybridization (see Supplemental Material~\cite{supp}). Consequently, we observe a marked enhancement in linewidths of acoustic and, much more evidently, the optical phonons at $\Gamma$ and M points beginning at 40 GPa [Fig.~\ref{fig3} (a)]. This is a consequence of intravalley and intervalley electron scatterings which are associated with phonons around $\Gamma$ and M, K, and H, respectively. For instance, considering the symmetry, twelve equivalent L points exist in the Brillouin zone [Fig.~\ref{fig4} (b)]. Each electron state at L can be linked to its nearest counterpart at L through a phonon at the M point (see Supplemental Material~\cite{supp}). A similar scenario unfolds around $\Gamma$, K and H phonon states. The increase in phonon linewidth at $\Gamma$ is associated with intravalley scattering between identical electron states located at different high symmetry points. Conversely, a slight reduction in phonon linewidth around K and H is linked to intervalley scattering between two electron states around A and K points, corresponding to the contraction of the Fermi surface area. However, the subsequent decrease in phonon linewidth due to phonon-electron interaction at the K and H points has less influence than the emergent phonon linewidth at M and $\Gamma$ points. Consequently, we observe a marked reduction in the slope of $\kappa_\texttt{ph}$ from 40 GPa when considering phonon-electron scattering rates [Fig.~\ref{fig2} (b)]. Note that only effects on the acoustic phonons are relevant to the thermal conduction, although the optical phonons have larger linewidths due to phonon-electron interaction than acoustic phonons do.

In summary, starting from first-principles calculations, we reveal a non-monotonic pressure dependence of $\kappa$ of TaN similar to that observed in BAs and BP but due to a different mechanism. This anomalous behaviour results from the competing responses of phonon-phonon and phonon-electron scattering to pressure, in contrast to the interplay between different phonon-phonon scattering channels in BAs and BP. The overall phonon dispersion stiffens almost at the same pace, leading to continuously weakened three-acoustic-phonon and four-phonon scatterings, giving rise to an initially increasing $\kappa$. At higher pressures, the increased electronic DOS, explicitly the emergence of additional pockets of the Fermi surface at the high-symmetry L point in the Brillouin zone, leading to an increase in phonon-electron scattering and consequently driving a decrease in $\kappa_{\mathrm{ph}}$. The $\kappa$ of TaN can surpass that of BAs in the range $\sim$~20$-$70~GPa. Finally, it is worth noting that TaN has been synthesized and remains stable at high pressures~\cite{Feng_CPB18,Jia_CPL19,Lee_AFM23}. Our reported anomalous pressure dependence of $\kappa$ occurs at around 60~GPa, and thus within the experimentally accessible range. The non-monotonic pressure dependence may also be observed in some other metals, where the electronic DOS changes suddenly with pressure.


The supporting data for this paper are available from Zenodo~\cite{Kundu_data_TaN} under open licenses. The data package includes the structure file, second-, third- and fourth-order IFCs and phonon-electron scattering rates for each pressure.

We acknowledge support from the National Key R\&D Program of China (Grant No. 2023YFA1407001), the Natural Science Foundation of China (NSFC) (Grants No. 12174261 and No.12104312), and the Guangdong Basic and Applied Basic Research Foundation (Grants No. 2021A1515010042, No. 2023A1515010365, and No. 2022A1515011877).  A. K. and M. K gratefully acknowledge the support and resources provided by the PARAM Brahma Facility at the IISER, Pune, under the National Supercomputing Mission of the Government of India. A. K. acknowledges CHANAKYA Post-Doctoral fellowship from the National Mission on Interdisciplinary Cyber-Physical Systems (NM-ICPS) of the Department of Science and Technology, Government of India, through the I-HUB Quantum Technology Foundation, Pune, India. X.Y. acknowledges support from the NSFC (Grants No. 12374038 and No. 12147102). 


%

\end{document}